# Analyzing Brain Circuits in Population Neuroscience: A case to be a Bayesian

Danilo Bzdok[1,2,3], Dorothea L. Floris[4], Andre F. Marquand[4,5]

[1] Department of Psychiatry, Psychotherapy, and Psychosomatics, Rheinisch-Westfälische Technische Hochschule (RWTH) Aachen University, Aachen, Germany

[2] Parietal Team, Institut National de Recherche en Informatique et en Automatique (INRIA),

[3] Neurospin, Commissariat à l'Energie Atomique (CEA) Saclay, Gif-sur-Yvette, France

[4] Department of Cognitive Neuroscience, Radboud University Medical Centre, Donders institute for Brain, Cognition and Behaviour, Nijmegen, Netherlands

[5] Department of Neuroimaging, Institute of Psychiatry, Psychology and Neuroscience, King's College London, U.K.

## Abstract

Functional connectivity fingerprints are among today's best choices to obtain a faithful sampling of an individual's brain and cognition in health and disease. Here we make a case for key advantages of analyzing such connectome profiles using Bayesian analysis strategies. They (i) afford full probability estimates of the studied neurocognitive phenomenon; (ii) provide analytical machinery to separate methodological uncertainty and biological variability in a coherent manner; (iii) usher towards avenues to go beyond classical null-hypothesis significance testing; and (iv) enable estimates of credibility around all model parameters at play and thus enable predictions with uncertainty intervals for single subjects. We pick research questions about autism spectrum disorder as a recurring theme to illustrate our methodological arguments.

## Introduction

In most fields of neuroscience, including connectomics, drawing statistical conclusions from brain data is essential to understand the measurements of the studied phenomenon despite the presence of noise. For example, to infer whether a given functional brain connection is strengthened or weakened by a certain stimulus or to predict a clinical diagnosis of a given subject on the basis of neuroimaging data. In this article, we argue that adopting a Bayesian perspective to such problems offers several benefits, which arise from the ability to coherently handle uncertainty in model predictions.

Bayesian analysis has a long history, with origins in the 18$^{th}$ century (Bayes, 1763). In essence, Bayesian statistics treats all parameters in a given model as random variables and quantifies their uncertainty using Bayes rule. More concretely, for a model $\mathcal{M}$ with parameters $\theta$ this rule indicates that we should update our *prior* belief in the probability of the parameters $p(\theta|\mathcal{M})$ in light of data, $y$, to derive the *posterior* distribution over the model parameters, $p(\theta|y,\mathcal{M})$.

$$p(\theta|y,\mathcal{M}) = \frac{p(y|\theta,\mathcal{M})p(\theta|\mathcal{M})}{p(\theta|\mathcal{M})}$$

The term $p(y|\theta,\mathcal{M})$ is the *likelihood* and provides a generative model that describes how the data may have come about. The denominator is referred to as the *marginal likelihood* and is obtained by integrating out the parameters $p(\theta|\mathcal{M}) = \int p(y|\theta,\mathcal{M})p(\theta|\mathcal{M})\,d\theta$. Posterior computation is intractable for all but the simplest models. This is why posterior parameter estimation often requires the use of numerical approximations or sampling methods, which is probably the major hurdle to the use of Bayesian methods in practice.

It is important to recognise that the Bayesian philosophy of data analysis assumes a deeper and more general notion of probability than is assumed by most of the quantitative tools commonly used in many areas of neuroscience. In particular, under the frequentist paradigm (Cox, 2006), probability reflects long run frequencies of repeatable events (e.g. 'the probability of rolling a 6 on this dice is 1/6' or 'the neural activity measured during a perspective-taking task'). Under the Bayesian paradigm, probabilities reflect degrees of belief in a given proposition (e.g. 'there is a low probability that the amygdala will activate 100 times more or less in autism vs. health'), which may not be repeatable. In a neuroscience context, frequentist notions are routinely used, especially for hypothesis testing against a null distribution, for example to define the probability that a given brain region shows more neural activity than would be expected under the null hypothesis of baseline activity.

Frequentist inference assumes that the data-generating mechanism is fixed and that only the data have a probabilistic component. Inference about the model is therefore indirect, quantifying the agreement between the observed data and the data generated by a putative model (for example, the null hypothesis). Instead, Bayesian inference quantifies the uncertainty about the data-generating mechanism by the prior distribution and updates it with the observed data to obtain the posterior distribution. Inference about the model is therefore obtained directly as a probability statement based on the posterior.

Direct quantification of uncertainty is the central goal of Bayesian analysis (Gelman et al., 2013). Bayesian modeling aims to coherently incorporate uncertainty throughout the analysis such that uncertainty in the parameter estimation is carefully propagated through the generative model to form predictions. It can be shown mathematically that probability theory is the only way that this can be achieved; on the basis of a simple and common-sense set of axioms (Cox, 1946; Jaynes, 2003). In short, any system of reasoning that coherently manages uncertainty must be consistent with the rules of probability. As such, when carrying out Bayesian data analysis, the analyst naturally goes beyond point estimates of parameters, such as a value indicating the association of the connectivity strengths between the amygdala and the prefrontal cortex. Rather than a single number, full probability distributions are placed over all quantities in the model which are updated in the light of the brain data at hand. Based on the build Bayesian model, new predictions can be formed for incoming data points by averaging (i.e. integrating over) the joint posterior distribution over all model variables on the table.

A defining characteristic of Bayesian analysis is that this regime requires the specification of a prior distribution, reflecting our beliefs about the model parameters before observing any data. Thus, each model parameter has a fully specified probability distribution, whether or not data have already been brought into play or not. If prior information is available (e.g. structural connectivity constraining functional connections, which functional connections describe intra-network versus between-network connections or whether connections between subcortical areas may be harder to measure than those between cortical areas), this can be included in the model. This helps guide the parameter updates to biologically plausible ranges in the face of new observations, while still permitting solutions that exceed the pre-set ranges to the extent supported by the data. Even in the absence of definite strong, biologically grounded a-priori information, generic priors can be employed to exert a regularizing or smoothing effect on the parameter estimation (e.g. to prevent overfitting connectomic profiles of a subject sample that may not extrapolate well to the broader population). However, the specification of the priors over model parameters is often a point of criticism for Bayesian methods, because it can often be difficult to specify informative priors if the number of variables are large or the dependencies between them are complex (Woolrich et al., 2009). Moreover, it is often not straightforward to specify priors that convey a lack of prior knowledge (Cox, 2006; Jeffreys, 1946). Nevertheless, it is important to recognise that all analysis methods are predicated on certain assumptions. The fact that the Bayesian approach forces these to be made explicit can be viewed as a strength.

**Notions of probability: Methodological uncertainty and biological variability**

An important distinction, often under-appreciated in neuroscience in the quite opinion of the authors, regards the type of probability that is actually being modelled. Under the Bayesian framework, probabilities can be used for multiple different purposes (Cox, 2006): they may be treated in a 'phenomenological' manner to quantify natural biological variation in the data (e.g. how different are amygdala-prefrontal connections across subjects in the population). However, probability can also be framed in an 'epistemological' manner to quantify uncertainty in parameter estimation (e.g. uncertainty in the estimation of amygdala volume due to finite data). This reflects the distinction in statistical machine learning (Gal, 2016; Kendall and Gal, 2017) between 'aleatoric' uncertainty which reflects inherent variation in the measured phenomenon in nature that cannot be reduced with acquiring more data and 'epistemic' uncertainty which reflects uncertainty in our knowledge of model parameters and data density can be reduced by adding more observations. Unfortunately, this nomenclature confounds the notions of variability and uncertainty described above. To simplify the discussion in this manuscript, we principally distinguish between (biological) *variability* and (methodological) *uncertainty*. For completeness, we note that in some cases it may be desirable to further decompose epistemic uncertainty (e.g. due to scanner noise or interpolation error).

Importantly, most dominant frequentist approaches conflate variability and uncertainty to a certain degree. Frequentist approaches – at best – provide post-hoc estimates of model *uncertainty* using techniques such as bootstrapping (Efron, 1979). For most applications, accurately quantifying *variability* is of primary interest, while minimizing or properly accounting for *uncertainty*. To provide a concrete example, normative modelling is a recently introduced technique that aims to map centiles of variation, such as the functional connectivity strength between amygdala and prefrontal cortex, across a reference cohort in an analogous manner to the use of growth charts in pediatric medicine (Marquand et al., 2019; Marquand et al., 2016a). For example, by plotting biological parameters as a function of age (or other clinically relevant variables), normative modelling enables statistical inferences as to where each individual participant falls within the population range. It can therefore be used to chart biological *variability* relevant to many disorders including autism and detect the biological signatures of brain disorders in an anomaly detection setting (Zabihi et al., 2018). In such applications, the primary interest is in modelling inter-subject *variation* across the cohort whilst accounting for modelling *uncertainty* such as noise intrinsic to the fMRI signal. For such neuroscience applications, the ability to jointly model different sources of variation and appreciate uncertainty in the same Bayesian model is an important advantage of the Bayesian approach. For example, using Bayesian methods the investigator can use separate variance components to model variation in age-related connection strength across a population cohort and the uncertainty in that estimation, due for example to data sampling density (e.g. fewer female subjects, or less high-functioning patients). In contrast, classical methods may also be used for normative modelling. Confidence intervals for the centiles of variation could be derived using bootstrapping (Huizinga et al., 2018). However, it is then difficult to include these estimates in subsequent statistical inferences; such as automatically detecting individuals with autism on the basis connectome fingerprints.

The value of Bayesian inference for delineating quantities of variability and uncertainty in neuroimaging connectivity analysis has been advertised through a substantial body of literature (e.g. Bowman et al., 2008; Friston et al., 2008; Friston et al., 2002a; Friston and Penny, 2003; Friston et al., 2002b; Penny et al., 2005; Woolrich, 2012; Woolrich et al., 2004; Woolrich et al., 2009). More specifically, in the context of brain networks, Bayesian methods have been applied for improving the estimation of whole-brain connectivity profiles (Colclough et al., 2018; Hinne et al., 2014) in finding parcellations of different brain networks (Janssen et al., 2015), for causal inference in fMRI (Mumford and Ramsey, 2014) and for multi-modal data fusion (Groves et al., 2011). These existing applications have largely focused on datasets of modest size, for which Bayesian methods are well-suited due to the regularizing effect exerted by the imposed priors and the guidance of parameter updates by existing neuroscience knowledge. For example, generic priors can be used to restrict the magnitude of model parameters, thus helping to prevent overfitting to idiosyncrasies of a particular subject sample of connectivity fingerprints, rather than being characteristic for the healthy or autism population more broadly. In addition to previous applications, we argue here that Bayesian methods also provide an excellent tool for large, population-based cohorts, which are gaining center stage in clinical neuroimaging (Di Martino et al., 2014; Miller et al., 2016b; Smith et al., 2015a; Van Essen et al., 2013; Volkow et al., 2017).

There are also several reasons for the suitability of Bayesian methodology in the 'big data' contexts (cf. Bzdok et al., 2019; Bzdok and Yeo, 2017): the ability to separately quantify variability in the natural phenomenon under study and uncertainty in the model under use is likely to be instrumental to understanding inter-individual variations across large cohorts (cf. above), the importance of which is increasingly recognised in IQ prediction based on connectivity fingerprints and other successful examples (Finn et al., 2015; Foulkes and Blakemore, 2018; Marquand et al., 2019; Seghier and Price, 2018). Bayesian methods are also appealing because they provide estimates of the plausible range of a parameter value given the data. In contrast, in large samples classical null-hypothesis testing methods can easily reject the null hypothesis for nearly all values (e.g. all voxels in a classical frequentist connection-wise analysis), even though the underlying effects are of negligible magnitude (see Friston and Penny, 2003). It is also obvious that quantifying methodological *uncertainty* is critical for optimal decision-making in medicine (Bishop, 2006). For example, for predicting an autism diagnosis on the basis of MRI scans, where uncertainty arises at multiple levels: not only in the diagnosis itself (i.e. at the level of clinical presentation), but also at the level of the underlying biology (e.g. the connectivity strength in a network modelling context).

In this paper, we will provide a conceptual overview of the aim and utility of Bayesian tools in clinical neuroscience, focusing principally on the use of such methods for connectomics. Functional connectivity fingerprints are particularly valuable for capturing salient characteristics of momentary states of conscious awareness and for predicting individual differences in cognition (Finn et al., 2015; Rosenberg et al., 2016; Smith et al., 2015a). These analytical techniques are widely applicable to predicting symptomatology across many clinical populations (Fornito et al., 2015; Xia et al., 2018). In the following, we highlight three particular application areas for Bayesian methods: (i) dealing with nuisance variation within the population; (ii) providing statistical uncertainty estimates for model parameters and predictions and (iii) stratification of clinical and population samples. For the sake of illustration, we provide concrete examples of these methods as applied to the study of autism spectrum disorder in large data cohorts.

**Hierarchical Bayesian modeling: Appreciating covariates of population stratification**

The boundary between signal and noise is often hard to identify; let alone to know prior to data analysis. It is common practice in many empirical sciences, including imaging neuroscience (Miller et al., 2016a; Smith et al., 2015b), to adjust for nuisance variance in the data in two separate steps. In a first step, variation that can be explained by nuisance covariates is removed, typically using linear-regression-based deconfounding. In a subsequent step, the remaining variation in the data is then fed into the actual statistical model of interest used to draw neuroscientific conclusions. As such, the final interpretation is typically grounded in model parameter estimates from a version of the original data, in which any linear association with the consider nuisance covariates, such as age- and sex-related differences between individuals, has been completely removed beforehand. In this scenario, the implicit but critical assumption is that any target effects of interest in the brain data, such as for the goal of classifying neurotypicals from individuals with a diagnosis of autism, is treated largely separately of what is measured by the nuisance covariates.

In many brain disorders including autism, the distinction between signal and noise may be more challenging to define. Age, sex, and motion are often chosen as nuisance covariates. However, the majority of autism samples includes 3-5 times more males who carry a diagnosis of autism than females (Kanner, 1943; Lai et al., 2017b; Scott et al., 2002), reflecting the prevalence in the wider population. While several reasons can be brought forward (Floris et al., 2018; Goldman, 2013; Lai et al., 2015; Schaafsma and Pfaff, 2014), it has been speculated that the discrepant prevalence of autism may point to a more profound distinction in the etiology of the disease, with its triggering life events, underlying pathophysiological mechanisms, and ensuing coping strategies. Preceding removal of sex-related signal in the data can also remove information of and preclude insight about sex-specific disease pathways in autism or lead to spurious or incorrect inferences (Miller and Chapman, 2001; Miller et al., 2016a). Let's take a hypothetical example where amygdala-prefrontal connectivity is pathologically increased in male patients, but pathologically decreased in female patients. Here, a preceding deconfounding step for sex would largely remove this sex-dependent aspect that truly is a characteristic of disease biology from subsequent statistical analysis and scientific conclusion.

Similarly, the age trajectories of male and female individuals with autism may be different in multiple ways. A commonly mentioned clinical feature of autism, for instance, is that females are more often only diagnosed later in life (Schaafsma and Pfaff, 2014). Better coping strategies and more successful camouflaging behavior in women with autism is a common explanation for this age-related divergence (Lai et al., 2017a). Consequently, removing age-related variance in brain measurements as a "data preprocessing" step can systematically withhold clues that can teach us something about the age-dependent development of autism in different strata of the population. Therefore, we argue that it is better to model shared variance explicitly, such as in jointly modeling age-dependent connectivity variation and autism-dependent variation in the functional connectome, for which Bayesian analysis is well suited. For example, Bayesian analysis could answer a question such as "How certain are we that a given brain connectivity feature is similar or dissimilar in certain subgroups, such as divided by sex or lifespan?".

Rather than taking a deterministic decision in a black-or-white fashion, Bayesian hierarchical modeling (BHM) is a natural opportunity to quantify the separate contributions by answering which sex-, age- and motion-related components in functional connectivity couplings are related to autism-related model parameters with which magnitude and how certain can the investigator be about it. A set of sources of variation in the brain data can be directly integrated in a single model estimation, instead of carrying out initial confound and later effect analyses (cf. above). Said in yet another way, BHM allows for explicit modeling of the group differences in functional brain connections in disease vs. control groups as linked to the question of how much any group difference are influenced by age, sex, and motion variation in the functional connectivity data by hierarchically accounting for dependencies between them. Removing age, sex and motion related information from the data in an isolated step hides important information that can be instrumental in guiding the actual model parameter estimation. While this goal can also be accommodated in a non-Bayesian setting (e.g. using linear mixed effects models), the Bayesian formulation is appealing because it coherently propagates uncertainty through different levels of the model and can therefore more readily disentangle different sources of variation and uncertainty.

We consider motion as another example of a covariate that is widely used to remove variation as measured by head movement during brain scanning. A few years ago, brain-imaging investigators have reported a seemingly distinctive patterns of maturing intrinsic brain fluctuations with successively weakening short-range and growing long-range connections that slowly change during child development (Dinstein et al., 2011; Dosenbach et al., 2010). Neuroscientists speculated these findings to mean that normal children start life with prominent short-range connectivity, which then weakens over the life span in healthy controls; vice versa for long-range connectivity (Belmonte et al., 2004). People with autism were then found to show more short-range and less long-range connectivity links (Geschwind and Levitt, 2007; Keary et al., 2009), especially in children. Unfortunately, it later became apparent that excessive head movements also reliably entailed artefacts with these same connectivity patterns in functional brain connectivity, previously thought to reflect impaired brain maturation (Power et al., 2012; Van Dijk et al., 2012), which entailed a series of retractions of high-profile papers ([weblink](weblink)).

On the other hand, at the behavioral level, it is well established that people with autism exhibit greater degrees of movement than healthy controls (Nordahl et al., 2008; Yendiki et al., 2014). As such, unusually high movement can be argued to be a hallmark feature of autism, but is now recognized to also be a reason for spurious functional connectivity findings. Put differently, it is hard to give a clear-cut answer which part of functional connectivity signals corresponds to motion-related noise and which part corresponds to neurobiologically informative signals in functional connectivity synchronization between brain regions. With and without a given adjustment for motion-related influences, distinguishing functional connectivity fingerprints in autism reflect different statistical questions (Pearl and Mackenzie, 2018). The data analysis scenarios correspond to two equally valid questions depending on the purpose. Adjustment relates to partitioning a population into groups that are homogenous according to the deconfounding variable - there may be no single right or wrong. Bayesian analysis can help in quantifying uncertainty via probability distributions over the extent motion measurements are related to brain connectivity strengths and to other measurements of interest in an integrated fashion.

The inferential granularity of conclusions about brain data can thus be enhanced by findings with hierarchical models that accommodate dependencies between parameters at different levels. In this way, BHM allows asking more ambitious questions using hierarchical population models of brain connectomics in strata of individuals. Young people with autism are different from old people with autism as reflected in their connectome profiles. An additional and not mutually exclusive source of variation is that male autism is different from female autism, conjointly across lifespan. We can estimate differences between autism and control groups by modeling hierarchical dependencies between multiple sources including covariates, like age, sex, and motion, with parameters corresponding to brain connectivity measurements. This multi-level setup allows for partial pooling of information between measurements suspect to exert confounding influence and genuine measurements of brain signals. For instance, neuroscientists may find that increased amygdala-prefrontal connectivity in autism is particularly characteristic for females who are in early childhood and tend to move their head little in the brain scanner as part of a joint posterior parameter distribution incorporating all measured sources of variability. Additionally, sex imbalance is often encountered in population samples of autism which can reflect the population prevalence or explicit exclusion of female cases. Imbalance in the considered participants in each group can be explicitly handled by BHM, with appropriate accounting for uncertainty. To adjust for these differences in naturally occurring group size we can avoid being misled in the way common single-level models typically would be (McElreath, 2015). As such the often made a-priori distinction into signal and noise, as a separate preprocessing step, can be relaxed by combining and integrating statistical evidence from disparate sources in a single probabilistic model estimation (Efron and Hastie, 2016).

## **The importance of saying no: Uncertainty estimates for single-subject predictions**

As one of various supporting hints for the biological basis of autism, the structure and function of the amygdala was repeatedly highlighted to differ in patients with autism, which is thought to play a role in impaired social interaction (Baron-Cohen et al., 2000). Statistically significant differences in the amygdala in autism led to varying reports in different patient samples (Kim et al., 2010; Nacewicz et al., 2006). Thus, this disease manifestation does not appear to be present in every single autism patient, nor to be consistently present on average in every patient sample recruited for studies that compare healthy and diagnosed individuals. Asking whether or not a strict categorical difference exists in a specific brain region in individuals on the autism spectrum may simply be a suboptimal analytical approach for the job.

Put differently, any statistical tool that is designed to give black-and-white categorical answers may be inappropriate for probing disease features that are a) present in autism patients to varying degrees (i.e., reflecting phenomenological *variability*), b) difficult to detect from the noisy behavioral and/or functional connectivity measurements that are available (i.e., reflecting epistemological *uncertainty*), or both. If these two sources of variation have played a role in amygdala studies in autism then using analysis approaches that can only make raw statements declaring presence or absence of an effect may be largely ill-suited. If the phenomenon under study is highly variable across people and/or tricky to quantify methodologically, then investigators in one lab may conclude on presence of a difference in connectivity between amygdala and prefrontal cortex on their sample, while another research group studying a different patient sample may converge on absence of group-related connectivity differences. While the answer is seemingly certain in each of these studies, the uncertainty in whether or not an amygdala effect is present in autism comes out at another end (Nosek et al., 2015): lacking reproducibility across different studies that have examined statistically significant (vs. insignificant) amygdala alterations in autism samples (He et al., 2019).

In such cases, the conventional frequentist 95% confidence intervals are not the solution that many investigators want: It is common to hear that a 95% confidence interval means that there is a probability 0.95 that the true parameter value lies within the interval; that is, that we do not have enough evidence to reject the null hypothesis of equal amygdala volume in both groups. In strict non-Bayesian statistical inference, such a statement is never correct, because strict non-Bayesian inference forbids using probability to measure uncertainty about parameters like a measure of amygdala volume in healthy vs. diseased individual. Instead, one should say that if we repeated the study and analysis a large number of different samples, then 95% of the computed intervals would contain the true parameter value. The classical 95% confidence interval only takes its meaning in the hypothetical long run of repeatedly analyzing new samples of controls and patients. Then we expect to be mistaken about presence of absence of amygdala effect in only 5%, that is 1 in 20 of the conducted neuroscience studies.

Rather than forcing definitive answers on presence against absence of subtle amygdala effects using null-hypothesis statistical significance testing, Bayesian analysis fully embraces unavoidable variation as an inherent part of model building, estimation, and interpretation (Gelman et al., 2014). In the Bayesian paradigm, each component of the model has a fully specified probability distribution (before and after seeing the data). As a consequence, a Bayesian model estimating amygdala differences in healthy vs. autistic individuals naturally provides estimates of the *degree of difference* at the phenomenological level as well as estimates of the modeling uncertainty at the epistemological level. Any amount of anatomical divergence between 0 ('no difference') and 1 ('difference') is a possible and legitimate result in the Bayesian posterior parameter estimate, while fully accounting for uncertainty in the parameter estimation.

In this way, the Bayesian modeling regime offers rigorous statements on how much a conclusion on a given group differences in the amygdala is justified in the patient sample at hand. The width of the corresponding parameter posterior estimate can be narrow to indicate high certainty in the obtained group difference. In contrast, the posterior distribution can be widely spread out to indicate low methodological certainty and thus limited neuroscientific trustworthiness of the found parameter value reflecting amygdala volume difference. Bringing together these different elements, Bayesian modeling directly provides a confidence judgment about each quantity on the table. For example, it allows statements such as: 'under the model, there is a 95% probability that amygdala volume differs between individuals with autism and controls'. If the evidence for the tested difference is ambiguous, we want this to be the result of the analysis so that we can align the strength of our conclusions with the certainty that the model provides.

Most common frequentist modeling approaches have a harder time telling the investigator when the modeling result is unsure or not. For example, linear support vector classifiers or linear discriminant analysis similarly vote for autism, rather than control, based on a brittle 51% or a solid 98% probability for evidence of group difference in the amygdala. In other words, Bayesian analysis frameworks are a rare opportunity where the resulting model solution "knows when it does not know". Moreover, in null-hypothesis statistical testing, the probability of detecting an effect (i.e. statistical power) increases with increasing sample size, even though the effect size (e.g. in terms of group differences in a point estimate for a given parameter) does not (Wagenmakers et al., 2008). Bayesian analysis does not suffer from that problem for the reasons we have outlined above.

The ability to say 'None' when the investigator asks for whether a group difference either exists or not will probably turn out to be crucial in our efforts towards precision medicine (Arbabshirani et al., 2017; Bzdok and Meyer-Lindenberg, 2018; Stephan et al., 2017). As Bayesian models are fully probabilistic by construction, brain data from a new incoming individual, such as brain scanning yielding amygdala volume measures, can be propagated through the already-built model into a *probabilistic prediction* for the single individual at hand. This additional information can be crucial in a variety of settings in neuroscientific research and clinical practice. First, generating single-subject predictions in a patient may yield different levels of certainty in assessments of autism symptoms related to language, motor behavior, IQ, or social interaction capacities. For example, individuals who are confidently classified may have more severe symptoms in a particular domain, whilst others that are less confidently classified may be more mildly affected. Separate judgments on the certainty of predictive conclusions in each of these symptom domains may turn out to characterize different types of autism in the spectrum, such as high-functioning autism. Second, along the life trajectory, different symptom dimensions of autism may turn out to be predictable based on brain measurements with higher or lower confidence, which may turn out to be characteristic for developmental periods in autism, or specific for atypical or different subtypes of autism. For instance, in women with autism, typically better camouflaging of social deficits (Dean et al., 2017; Hull et al., 2017; Lai et al., 2017a) may lead to social impairment predictions that have non-identical confidence in men with autism. Third, uncertainty is undoubtedly a key asset of treatment response prediction to choose therapeutic interventions tailored to single individuals. In this context, models predicting which treatment option to choose will be all the more useful in clinical practice, if such algorithmic recommendations also carry forward information on the forecasting confidence.

## Disease subtyping: Towards probabilistic intermediate phenotype discovery

A key challenge in the study of most psychiatric disorders – including autism – is that individuals with the same clinical diagnosis vary considerably from one another in terms of clinical phenotype and underlying neurobiology. This has led to some proposing that it may be preferable to consider the 'autisms' (Geschwind and Levitt, 2007). Many studies have aimed to dissect the clinical phenotype of autism (Wolfers et al., 2019b), for which functional connectivity provides promising candidate features (e.g. (Easson et al., 2019)). Moreover, since atypicalities are often complex and multifaceted, the features used for this purpose are often high-dimensional (e.g. consider even more whole-brain nodes in functional connectivity matrices) and/or multimodal (e.g. combine measures derived from structural and functional connectivity).

In general, the goal of such studies is to infer the latent structure underlying the clinical phenotype (e.g. partitioning the cohort into subtypes) on the basis of psychometric or biological variables, whilst accounting for nuisance variation. There are many ways that this can be achieved, including classical clustering techniques and matrix factorization techniques such as non-negative matrix factorization (NMF) and independent components analysis (ICA). Briefly, clustering approaches focus on finding subtype clusters in the data, whereas matrix factorization approach focus on finding useful decompositions of a data matrix under various assumptions. This can be used, for example, to find latent factors that may overlap across individuals in that any given individual may express multiple latent factors (Tang et al., 2019). Whilst these approaches are widely applied in a classical frequentist context, Bayesian variants have also been developed. In addition, highly promising Bayesian 'non-parametric' clustering and matrix factorization approaches have been developed such as Dirichlet process mixtures ('Chinese restaurant processes') and the 'Indian buffet' process (IBP) . Adopting a Bayesian approach to such problems confers many benefits, including providing good control over latent representations of the data, thereby helping to attenuate problems with high-dimensional estimation, providing predictive intervals around parameter estimates and predictions and providing flexible noise models for different forms of data. Moreover, Bayesian models are always generative in the sense that they always provide a model for how the data may have been generated. Collectively such Bayesian approaches are increasingly applied in clinical and neuroimaging contexts (Groves et al., 2011; Janssen et al., 2015; Ruiz et al., 2014; Schmidt et al., 2009)

A key problem in most classical stratification techniques is the issue of model order selection, or in other words, determining the optimal number of clusters or latent factors for the data at hand (Eickhoff et al., 2015). For example, "How many subtypes of autism exist in a given clinical dataset?". This is a notoriously difficult problem in classical statistics for which no uniquely optimal solution has imposed itself (Bzdok, 2017; Kleinberg, 2002), leading to suggestions that model order selection is perhaps sometimes largely a matter of taste (Hastie et al., 2009). There are many heuristic approaches for this problem, but these are subject to difficulty in practice. Additionally, choosing between a variety of viable model order selection criteria, which are by themselves objective, still amounts to taking a subjective choice on the number of latent factors best supported by the data. Different *cluster validity criteria* often give different answers or do not indicate a clear preference for one model order over others (Shalev-Shwartz and Ben-David, 2014), nor whether a given clustering solution explains the data better than a continuous model (i.e. with no clusters (Liu et al., 2008)). This has contributed to inconsistencies in the clinical stratification literature such that there are no consistently reported subtypes for autism (Wolfers et al., 2019a) or indeed any psychiatric disorder, despite decades of effort (Marquand et al., 2016b).

Bayesian non-parametric approaches provide an appealing solution to this problem because they can automatically adjust the model complexity (e.g. number of clusters or latent factors) on the basis of the data at hand. In other words, non-parametric models allow the flexibility of the model to grow with the number of data points used for model building. The simplest examples of Bayesian non-parametric models are Gaussian process models (Rasmussen and Williams, 2006), which are widely used for non-linear regression and have been used in normative modelling approaches described above. In a similar manner, Dirichlet process mixtures (DPM) (Ferguson, 1973; Neal, 1992) and 'Indian buffet' processes (IBP) (Griffiths and Ghahramani, 2011) provide an elegant potential solution to the problem of model order selection in clustering and matrix factorization, respectively. For example, the DPM model can be viewed as a clustering model with where the number of clusters is bounded only by the sample size, effectively making the DPM an infinite mixture model (Rasmussen, 2000). This has already been shown to be useful in a recent neuroimaging study on autism (Kernbach et al., 2018). As noted, a very appealing feature of this model is that it is self-calibrating in that it allows the optimal model order (i.e. the number of clusters) to be automatically inferred from the data whilst allowing the model order to grow with more data (i.e. increasing representational capacity). In practice, the number of clusters often grows sub-linearly with the number of observations (Neal, 1992). At the same time, by computing (or approximating) the full posterior distribution over the model parameters, this approach helps to attenuate overfitting. This non-parametric clustering approach has clearly desirable features for the stratification of psychiatric disorders such as autism in large data cohorts. Particularly as the size of the available datasets grows (e.g. through larger consortia), such models offer the ability to offer increasingly more fine-grained fractionations of the clinical phenotype. Similarly, in the context of brain networks, this approach has been shown to be useful for automatically parcellating brain networks into component regions (Janssen et al., 2015).

A more recent addition to the Bayesian non-parametric toolbox is the IBP (Griffiths and Ghahramani, 2011) is. The name is derived by analogy to the 'Chinese restaurant process' formulation of Dirichlet process mixtures (see (Aldous, 1985)). The IBP differs in that it does not assume that a single class is responsible for generating each data point (i.e. it does not provide a hard clustering solution). Rather, it allows each data point to express multiple features simultaneously, potentially reflecting multiple causes. Whilst this approach is yet to see extensive applications in brain connectomics, IBPs have been shown to provide an elegant way to model comorbidity in psychiatric disorders, where each individual expresses multiple latent factors to varying degrees (Ruiz et al., 2014).

The key advantage of Bayesian techniques for model order selection in contexts such as this is that they provide a formal framework for reasoning over model structures and inferring the plausibility of different candidate structures in view of the data at hand (see e.g. (Ghahramani, 2015; Tenenbaum et al., 2011). More generally, in cases where it may be computationally difficult to integrate out all model parameters, Bayesian models can also be used to infer optimal model parameters, an approach generally referred to as 'empirical Bayes'.

## Conclusion

In this concept paper, we provided a primer of Bayesian techniques for modeling brain circuits in health and their disturbance in autism spectrum disorder. Clearly, there are many different ways in which adopting a Bayesian approach to data analysis can open new windows of interpretations for neuroscience investigators.

We highlighted four special features of Bayesian methods: these analysis tools (i) provide an appealing interpretation of probability in terms of degrees of belief in a proposition, which is more general than the more restricted notion of reasoning about long-run frequencies of repeatable events; (ii) provide analytical machinery to separate (methodological) uncertainty and (biological) variability along with a calculus for reasoning about both in a coherent manner; (iii) usher towards avenues away from classical null-hypothesis significance testing, which is particularly valuable in data richness, and may contribute to overcoming the current reproducibility crisis in biomedicine. Finally, Bayesian methods (iv) afford estimates of uncertainty around all model parameters at play and can hence form predictions about single individuals by appropriate handling of all considered sources of variation.

An important practical limitation of Bayesian models is computational scalability. Since it is usually necessary to compute high-dimensional integrals to infer probability distributions for all quantities of interest, inference can be highly demanding in terms of memory or computation time or both. Markov Chain Monte Carlo (MCMC) methods (Neal, 1996) have made tremendous progress of recent years, and are now widely applicable to a point where these procedures for model estimation are regarded as the 'gold standard'. Generic sampling techniques methods are also now implemented in widely available software packages (Carpenter et al., 2017; Salvatier et al., 2016). This ready access makes Bayesian tools easier to deploy for a wider set of research labs and scientific questions than was previously the case.

We anticipate that in the coming years, Bayesian methods in general, and their hierarchical variants in particular, will be increasingly endorsed for their value in many applications of studying brain network connectivity, including data fusion, individual prediction, subgroup stratification of cohorts and for precisely quantifying statistical differences between experimental cohorts.